\documentclass[pra,aps,amsfonts,amsmath,twocolumn,showpacs,floatfix,superscriptaddress,nofootinbib]{revtex4}
\usepackage{bm}
\usepackage{amsfonts}
\usepackage{amsmath}
\usepackage{graphicx}
\newcommand{\vek}[1]{\bm{\mathrm{#1}}}
\newcommand{\up}{\uparrow}
\newcommand{\down}{\downarrow}
\newcommand{\Fig}[1]{Fig.~\ref{#1}}

\newcommand{\Secs}[1]{Secs.~\ref{#1}}

\newcommand{\Ref}[1]{Ref.~\cite{#1}}
\newcommand{\Eq}[1]{Eq.~(\ref{#1})}
\newcommand{\Eqs}[1]{Eqs.~(\ref{#1})}
\newcommand{\OmegaF}{\Omega_F}
\newcommand{\odd}{\text{odd}}
\newcommand{\hh}{\text{hh}}

\newcommand{\particle}{\text{p}}
\newcommand{\hole}{\text{h}}
\newcommand{\pp}{\text{pp}}
\newcommand{\pv}{\vek{p}}
\newcommand{\kv}{\vek{k}}

\newcommand{\dthreek}{\frac{d^3k}{(2\pi)^3}}
\newcommand{\dthreep}{\frac{d^3p}{(2\pi)^3}}
\newcommand{\dthreepp}{\frac{d^3p'}{(2\pi)^3}}

\newcommand{\calE}{\mathcal{E}}
\DeclareMathOperator{\Imag}{Im}

\DeclareMathOperator{\sgn}{sgn}

\begin{document}
\title{Occupation numbers in strongly polarized Fermi gases and the
  Luttinger theorem}
\author{Michael Urban}
\affiliation{Institut de Physique Nucl{\'e}aire, CNRS-IN2P3 and
Universit\'e Paris-Sud, 91406 Orsay Cedex, France}
\author{Peter Schuck}
\affiliation{Institut de Physique Nucl{\'e}aire, CNRS-IN2P3 and
Universit\'e Paris-Sud, 91406 Orsay Cedex, France}
\affiliation{Laboratoire de Physique et Mod\'elisation des Milieux
Condens\'es, CNRS and Universit\'e Joseph Fourier, Maison des
Magist\`eres, BP 166, 38042 Grenoble Cedex, France}
\begin{abstract}
We study a two-component Fermi gas that is so strongly polarized that
it remains normal fluid at zero temperature. We calculate the
occupation numbers within the particle-particle random-phase
approximation, which is similar to the Nozi\`eres-Schmitt-Rink
approach. We show that the Luttinger theorem is fulfilled in this
approach. We also study the change of the chemical potentials which
allows us to extract, in the limit of extreme polarization, the
polaron energy.
\end{abstract}
\pacs{03.75.Ss,67.85.Lm}
\maketitle
\section{Nozi\`eres-Schmitt-Rink approach and the particle-particle 
random-phase approximation}
Already many years ago, the possibility of a cross-over from Cooper
pairs in the Bardeen-Cooper-Schrieffer (BCS) state to a Bose-Einstein
condensate (BEC) of molecules was discussed theoretically
\cite{Eagles,Leggett}. Later, Nozi\`eres and Schmitt-Rink (NSR)
developed a theory that correctly interpolates between the critical
temperatures in the two limits \cite{NSR}. In this theory, only equal
densities of the two species forming the pairs (which we will denote
by spin indices $\sigma = \up,\down$) were considered. Nowadays, the
crossover can be realized in experiments with ultracold trapped atoms
whose scattering length $a$ can be tuned with the help of a Feshbach
resonance \cite{Greiner}. In these experiments, it is also possible to
study systems with different densities of the two species, $n_\up \neq
n_\down$ \cite{Shin}. In this way, one tries to discover, e.g., phases
with exotic pairing, like the Fulde-Ferrel-Larkin-Ovchinnikov (FFLO)
phase \cite{FF,LO}. Also the extremely polarized case, where only a
single particle of spin $\down$ is put into a Fermi sea of spin $\up$,
bears interesting physics: when passing though the Feshbach resonance
from the attractive ($a<0$) to the repulsive ($a>0$) side, one expects
that the ground state transforms from a Fermi sea plus a fermionic
quasiparticle, the so-called polaron, into a Fermi sea plus a bosonic
molecule \cite{Punk}. To our knowledge, there is so far no unique
many-body theory that can describe the imbalanced Fermi gas in the
cross-over and that reproduces in the limit of extreme polarization
the polaronic and the molecular ground state, depending on the value
of the scattering length $a$.

\begin{figure}
\begin{center}
\includegraphics[scale=0.5]{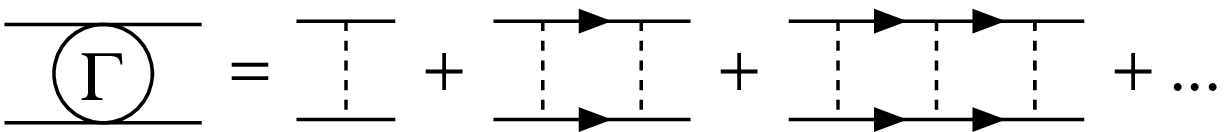}
\caption{Feynman diagrams for the T matrix in ladder
  approximation. \label{figladder}}
\end{center}
\end{figure}

\begin{figure}
\begin{center}
\includegraphics[scale=0.5]{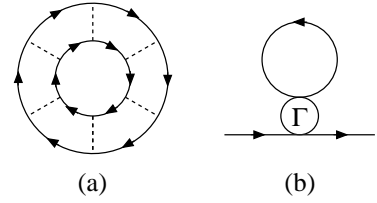}
\caption{(a) Typical diagram for the thermodynamic potential in the
  NSR approach. (b) Self-energy included to first order in the
  calculation of the correlated density. \label{figthermopot}}
\end{center}
\end{figure}

As mentioned before, the original NSR theory was formulated in order
to describe the BEC-BCS crossover in a two-component ($\sigma =
\up,\down$) Fermi gas with equal populations. Within this approach,
the critical temperature $T_c$ as a function of the chemical potential
$\mu$ is obtained from the Thouless criterion, i.e., it is the
temperature where the in-medium T matrix develops a pole,
\begin{equation}
\Gamma^{-1}(\omega=0,\kv=0) = 0\,,
\label{eqThouless}
\end{equation}
where $\omega$ and $\kv$ are, respectively, the total energy (measured
from $2\mu$) and momentum of the pair, and $\Gamma$ is obtained by
summing ladder diagrams, see \Fig{figladder}.
As a function of $\mu$, the critical temperature obtained in this way
is exactly the same as within BCS theory. The difference between BCS
and NSR comes from the inclusion of pair correlations into the
relationship $n(\mu)$ between the number density and the chemical
potential. This is done by including diagrams of the type shown in
\Fig{figthermopot}(a)
into the thermodynamic potential $\Omega(\mu,T)$ and then computing
the density from $n = -\partial\Omega/\partial\mu$. This is
equivalent to calculating the density from \cite{FetterWalecka}
\begin{equation}
n = 2T \int \dthreep \sum_{n~\odd} e^{i\omega_n \eta} G(i\omega_n,\pv)\,,
\label{eqnMatsubara}
\end{equation}
where $\omega_n = n \pi T$ is a fermionic ($n$ odd) Matsubara
frequency and $G$ is the single-particle (s.p.) Green's function with
at most one self-energy insertion,
\begin{equation}
G = G_0 + G_0^2 \Sigma\,,
\label{eqnfirstorder}
\end{equation}
as displayed in \Fig{figthermopot}(b). Notice that in the literature
one can also find variants of the NSR scheme where the Dyson series
for $G$ is summed to all orders, e.g., \Ref{Perali2002}.

Before turning to the imbalanced case, we want to discuss in some
detail the relation between the NSR scheme and the random-phase
approximation (RPA), here in the so-called particle-particle (pp)
channel (pp-RPA) \cite{RingSchuck} in contrast to the more familiar
particle-hole (ph) channel (ph-RPA). While the latter consists in the
resummation of ``bubble diagrams'', the pp-RPA consists in a
resummation of ladder diagrams as shown in \Fig{figladder}. In both
ph- and pp-RPA, the lines correspond to the propagators obtained at
the Hartree-Fock (HF) level. At $T=0$, which is the case we consider
in the present work, the propagators are chronological ones and belong
to a fixed Fermi momentum $k_F$ and not to a fixed chemical potential
$\mu$. The RPA correlation energy (i.e., correction to the HF energy)
can be obtained from the usual coupling-constant integration
\cite{FetterWalecka}, where one integrates, however, only over the
coupling constant appearing explicitly in the bubble or ladder
diagrams, respectively, keeping the HF field fixed. The formula
(\ref{eqnfirstorder}) for the Green's function to be used, e.g., in
the calculation of the correlation energy has to be slightly modified:
now $G_0$ denotes the HF Green's function, and $\Sigma$ is the
self-energy without the HF field.

In practice, the subtleties about whether one has to use HF or free
propagators are not relevant for us, since in the case of a
renormalized zero-range interaction, as it is generally used in
ultracold atom systems, the HF shift vanishes anyway
\cite{Perali2002}. However, the fact that we work with
zero-temperature propagators corresponding to a fixed density and not
to a fixed chemical potential is very important. Among other things,
it ensures that the pp-RPA formalism satisfies the Luttinger
theorem \cite{Luttinger}.

It is straight-forward to extend the NSR theory to the imbalanced
case, by introducing two different chemical potentials $\mu_\sigma$
($\sigma = \up,\down$). Unfortunately, as it was already observed by
several authors \cite{LiuHu06,ParishMarchetti07,KashimuraWatanabe12},
this scheme that works nicely in the balanced case fails in the
imbalanced case. To be specific, the problem is that near the unitary
limit (i.e., for large scattering length: $|a|\to\infty$) one
finds in some regions of the phase diagram $\rho_\up < \rho_\down$ in
spite of $\mu_\up > \mu_\down$.

We suspect that this problem is related to the fact that within the
NSR scheme the undressed s.p. Green's functions used to build the
ladder diagrams are computed with the same chemical potentials as the
corrected Green's functions. In the $T\to 0$ limit this implies that
the ladders are calculated in a system whose Fermi momenta are
different from the final ones. As it will be shown in a separate
article \cite{PA}, a finite-temperature formalism that includes the
shift of the s.p. energies self-consistently does not present the
pathological behavior of the NSR scheme and reduces to the pp-RPA in
the $T\to 0$ limit. The aim of the present paper is to see what
happens at $T=0$ within pp-RPA in the strongly imbalanced case.

\section{Particle-particle RPA for the strongly imbalanced case at
  zero temperature}
We consider now a polarized Fermi gas in which the density of $\up$
particles is higher than that of $\down$ particles, $n_\up >
n_\down$. At very strong polarization $P =
(n_\up-n_\down)/(n_\up+n_\down)$, the system remains normal fluid even
at zero temperature. It is this case that we want to discuss now. This
case includes in particular the polaron, i.e., a single $\down$
particle in a Fermi sea of $\up$ particles, which has recently
attracted a lot of attention from theoretical and experimental side
\cite{Chevy}.

At zero temperature, it is not necessary to use the Matsubara
formalism. Instead, one can start from the usual time-ordered
s.p. Green's function \cite{FetterWalecka}
\begin{equation}
G_0^\sigma(\omega,\pv) =
  \frac{\theta(k_F^\sigma-p)}{\omega-\epsilon_{\pv}-i\eta}
  +\frac{\theta(p-k_F^\sigma)}{\omega-\epsilon_{\pv}+i\eta}\,,
\label{eqnGzero}
\end{equation}
where $k_F^\sigma$ denotes the Fermi momentum of the atoms in spin
state $\sigma$. In contrast to the finite-temperature case discussed
before, the s.p. energies are in this formalism not measured from the
respective Fermi surfaces, i.e., $\epsilon_{\pv} = p^2/(2m)$
(throughout the article we use units with $\hbar = 1$, $\hbar$ being
the reduced Planck constant).

We mention that this formalism is not equivalent to the
zero-temperature limit of the Matsubara formalism: on the one hand,
within the Matsubara formalism, the Green's function $G$ is expressed
in terms of free Green's functions $G_0$ corresponding to the same
chemical potentials. On the other hand, in the zero-temperature
formalism, $G$ is expressed in terms of $G_0$ corresponding to the
same densities. The importance of this subtlety in the case of
non-perturbative resummations (such as ladder diagrams) will become
clearer below.
\subsection{In-medium T matrix \label{sectmatrix}}
Let us start by calculating the in-medium T matrix shown in
\Fig{figladder} within the zero-temperature formalism. As interaction,
we consider a contact interaction with coupling constant $g < 0$. Then
the T matrix can be written as
\begin{equation}
 \Gamma(\omega,\kv) =
   \frac{1}{\frac{1}{g}-J(\omega,\kv)}\,,
\end{equation}
with $J(\omega,\kv) = J_{\hh}(\omega,\kv) + J_{\pp}(\omega,\kv)$ and
\begin{align}
J_{\hh}(\omega,\kv) =& 
-\int^\Lambda \dthreep \frac{\theta(k_F^\up-p)\theta(k_F^\down-|\kv-\pv|)}
    {\omega-\epsilon_{\pv}-\epsilon_{\kv-\pv}-i\eta}\,,
  \label{eqnJhh}\\
J_{\pp}(\omega,\kv) =& \int^\Lambda \dthreep
  \frac{\theta(p-k_F^\up)\theta(|\kv-\pv|-k_F^\down)}
    {\omega-\epsilon_{\pv}-\epsilon_{\kv-\pv}+i\eta}\,.
  \label{eqnJppcutoff}
\end{align}
The subscripts hh and pp denote the contributions of hole-hole and
particle-particle propagation, respectively. The cutoff $\Lambda$ has
been introduced because the momentum integral in $J_{\pp}$ diverges.
To be precise, the integration region is defined by
$|\pv-\kv/2|<\Lambda$, and the upper integration limit in \Eqs{eqnJhh}
and (\ref{eqnJppcutoff}) should only be interpreted as a short-hand
notation. The usual procedure to deal with the divergence consists in
making the coupling constant $g$ dependent on $\Lambda$ and then
taking the limit $\Lambda\to\infty$, keeping the scattering length $a$
constant (which implies $g\to 0$) \cite{Perali2002}. The result of
this renormalization procedure can be written as follows:
\begin{equation}
 \tilde{\Gamma}(\omega,\kv) =
 \frac{1}{\frac{1}{\tilde{g}}-\tilde{J}(\omega,\kv)}\,,
\end{equation}
where $\tilde{g} = 4\pi a/m$,
$\tilde{J} = \tilde{J}_{\hh}+\tilde{J}_{\pp}$,
$\tilde{J}_{\hh}=\lim_{\Lambda\to\infty} J_{\hh}$, and
\begin{equation}
\tilde{J}_{\pp}(\omega,\kv) = \int \dthreep \Big(
  \frac{\theta(p-k_F^\up)\theta(|\kv-\pv|-k_F^\down)}
    {\omega-\epsilon_{\pv}-\epsilon_{\kv-\pv}+i\eta}+\frac{m}{p^2}\Big)\,.
  \label{eqnJppreg}
\end{equation}
The integrals $\tilde{J}_{\hh}$ and $\tilde{J}_{\pp}$ can be evaluated
analytically.

Note that within the RPA scheme we should in principle have started
from the HF Green's function instead of the non-interacting one
\cite{RingSchuck}. Consequently, the s.p. energies $\epsilon_{\pv}$ in
\Eqs{eqnJhh} -- (\ref{eqnJppreg}) should be replaced by HF energies
$\epsilon_{\pv}^\sigma = p^2/(2m)+gn_{\bar{\sigma}}$, where
$\bar{\sigma}$ denotes the spin opposite to $\sigma$. However, as
mentioned before, since $g\to 0$ in the limit $\Lambda\to\infty$, we
shall not bother with this unnecessary complication.

\begin{figure}
\begin{center}
\includegraphics[scale=0.5]{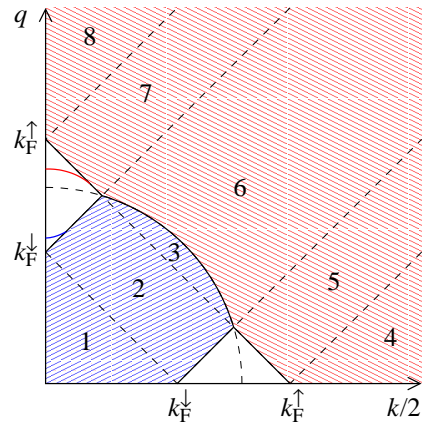}
\end{center}
\caption{Regions corresponding to different cases in the calculation
  of the function $\tilde{J}(\omega,\kv)$ in terms of the momentum in
  the center-of-mass frame, $q = \sqrt{m\omega-k^2/4}$, and the total
  momentum of the pair, $k$. The circle corresponds to $\omega =
  \OmegaF = \epsilon_F^\up+\epsilon_F^\down$ and separates the hh
  continuum (regions 1--3) from the pp continuum (regions 4--8). The
  solid lines in the triangle between regions 2 and 7 show
  schematically the positions of the poles $\Omega_{1,2}$ of
  $\tilde{\Gamma}$. \label{figregions}}
\end{figure}

\begin{figure}
\begin{center}
\includegraphics[scale=1.2]{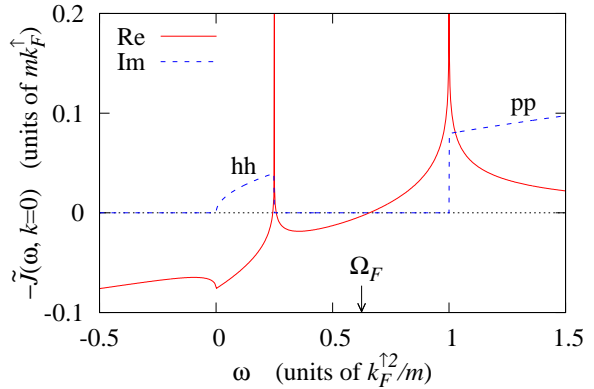}
\end{center}
\caption{Real (solid line) and imaginary (dashed line) parts of
  $-\tilde{J}(\omega,\kv=0)$ as function of $\omega$ for $k_F^\down =
  k_F^\up/2$ (corresponding to a polarization of $P \approx
  0.78$).\label{figplotj0}}
\end{figure}

It is useful to analyze in more detail the properties of $\tilde{J}$
and $\tilde{\Gamma}$. We define the variable $q$ corresponding to the
on-shell momentum of each atom in the center of mass (c.m.) frame of
the pair, via $\omega = q^2/m+k^2/(4m)$. In \Fig{figregions},
we show schematically the regions where the imaginary part of
$\tilde{\Gamma}$ is non-zero. The circle corresponds to $\omega =
\OmegaF = \epsilon^\up_F + \epsilon^\down_F$, where $\epsilon_F^\sigma
= k_F^{\sigma 2}/(2m)$. This circle separates the regions (1) to (3)
(hatched in blue), where the imaginary part comes from
$\tilde{J}_\hh$, from the regions (4) to (8) (hatched in red), where
the imaginary part comes from $\tilde{J}_\pp$. On the dashed lines
separating the different regions, $\tilde{J}$ has cusps. We see that
for $k < k_F^\up-k_F^\down$ the hh and pp continua are separated by a
region around $\omega = \OmegaF$ where the imaginary part
vanishes. This is also visible in \Fig{figplotj0}, where we display
the real and imaginary parts of $-\tilde{J}$ for the case $k_F^\down =
k_F^\up/2$ and $k=0$.
The sharp edges of the hh and pp continua lead to logarithmic
singularities in the real part of $\tilde{J}$. Similarly to the
original Cooper problem \cite{Cooper}, a small attractive interaction
($a < 0$) leads therefore to the existence of two poles in
$\tilde{\Gamma}$ (at the energies where $-\tilde{J}$ crosses
  $-1/\tilde{g}$): one slightly below the edge of the pp continuum,
corresponding to a bound pair of two particles, and the other slightly
above the edge of the hh continuum, corresponding to a bound pair of
two holes. These states are shown schematically as the red and blue
lines in \Fig{figregions}. With increasing total momentum $k$, the
sharp edges are washed out [regions (2) and (7) in \Fig{figregions}]
and the poles disappear.

As the interaction strength increases, the upper (pp) pole is shifted
to lower and lower energy until it reaches $\omega=\OmegaF$. As in the
NSR case, this indicates the onset of a pairing instability. If one
further increases the interaction strength, the pp pole first enters
into the energy range $\omega < \OmegaF$ of hh excitations, and then,
once $1/\tilde{g}$ drops below the minimum of $-\tilde{J}$, the pole
leaves the real $\omega$ axis and becomes complex.

As noticed in \Ref{LiuHu06} in the NSR framework, it is in the
imbalanced case not sufficient to consider only $\kv=0$ in the
Thouless criterion (\ref{eqThouless}), but the critical temperature is
the highest temperature where the T matrix has a pole at
$\omega=\OmegaF$ for any value of $\kv$, related to a transition
towards a FFLO-like phase with oscillating order parameter. In the
present case we are not interested in the critical temperature (since
we are at $T=0$), but in the critical polarization. If we start with a
fully polarized system ($P=1$) and decrease the polarization, the
critical polarization $P_c$ is reached when the T matrix has for the
first time a pole at $\omega=\OmegaF$ for any value of $\kv$. Actually
it is enough to check that
\begin{equation}
\tilde{J}(\OmegaF,\kv) > \frac{1}{\tilde{g}}
\label{conditionabovecritical}
\end{equation}
is fulfilled for all $\kv$. 

Experiments in the unitary limit \cite{Shin}, however, show that the
transition from the unpaired to the paired phase at low temperature is
not of second, but of first order, leading to phase separation between
unpaired and paired phases. Our theory does not allow us to check
whether a first-order transition appears already at higher
polarization than our $P_c$, since this requires a calculation of the
energy of the system in the paired phase.

\subsection{Self-energy and occupation numbers}\label{secselfenergy}
Let us start by writing down the expression for the self-energy
diagram shown in \Fig{figthermopot}(b),
\begin{multline}
\Sigma^\sigma(\omega,\pv) 
  = -i \int^\Lambda \dthreepp \int \frac{d\omega'}{2\pi}
    G_0^{\bar{\sigma}}(\omega',\pv')\\
    \times \Gamma(\omega+\omega',\pv+\pv')\,.
  \label{eqnSigma}
\end{multline}
For formal derivations, it will be sometimes convenient to keep the
cutoff finite (here, the integration region is defined by
$|\pv-\pv'|/2<\Lambda$), but in all practical calculations we will use
the renormalization procedure, i.e., replace $\Gamma$ by
$\tilde{\Gamma}$ and let the cutoff $\Lambda$ go to infinity.

\begin{figure}
\begin{center}
\includegraphics[scale=0.5]{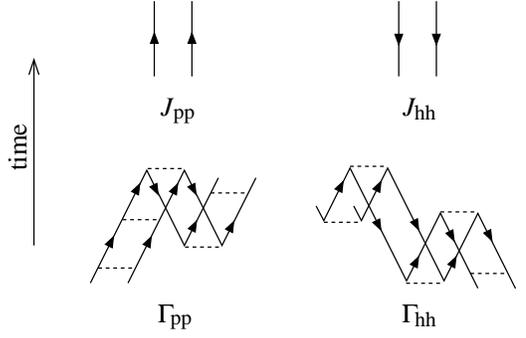}
\caption{Diagrammatic illustration of forward and backward going
  parts: free pp and hh propagators $J_{\pp}$ and $J_{\hh}$ (top),
  typical contributions to the forward and backward going parts of the
  T matrix, $\Gamma_{\pp}$ and $\Gamma_{\hh}$ (bottom).
  \label{figforwardbackward}}
\end{center}
\end{figure}

To evaluate \Eq{eqnSigma}, it is helpful to split $\Gamma$ into the
bare interaction $g$ (which vanishes in the limit $\Lambda\to\infty$)
and forward and backward going parts,
\begin{equation}
\Gamma = g + \Gamma_{\pp} + \Gamma_{\hh}.
\label{eqforwardbackward}
\end{equation}
Note that in $\Gamma$ the pp and hh channels are summed up together,
so that $\Gamma_{\pp}$ contains also contributions from hh propagation
and vice versa, as illustrated in \Fig{figforwardbackward}. 
The separation into forward and backward going parts can be achieved
with the help of dispersion relations:
\begin{align}
\Gamma_{\pp}(\omega,\kv) =& -\int_{\OmegaF}^\infty\frac{d\omega'}{\pi} 
  \frac{\Imag \Gamma(\omega',\kv)}{\omega-\omega'+i\eta}\,,
\label{eqnGammapp}\\
\Gamma_{\hh}(\omega,\kv) =& \int_{-\infty}^{\OmegaF}\frac{d\omega'}{\pi} 
  \frac{\Imag \Gamma(\omega',\kv)}{\omega-\omega'-i\eta}\,.
\label{eqnGammahh}
\end{align}
The imaginary parts in the numerators are meant to include also the
contribution of possible poles of $\Gamma$ at $\omega = \Omega_i$,
i.e.,
\begin{equation}
\Imag \Gamma(\omega,\kv) = \frac{\Imag J}{|\frac{1}{g}-J|^2} -\pi
\sum_i S_i(\kv)\delta\big(\omega-\Omega_i(\kv)\big)\,,
\label{eqnGammaspectrum}
\end{equation}
where the strengths of the poles are given by $S_i =
1/(dJ/d\omega)_{\Omega_i}\sgn(\OmegaF-\Omega_i)$. The self-energy can
now also be written as a sum of the energy-independent HF term and
forward and backward going parts,
\begin{equation}
\Sigma^\sigma = g n_{\bar{\sigma}} +
\Sigma^\sigma_{\pp}+\Sigma^\sigma_{\hh}\,,
\label{eqnSigmasplit}
\end{equation}
with
\begin{align}
\Sigma^\sigma_{\pp}(\omega,\pv) 
   =& \int^\Lambda \dthreepp \theta(k_F^{\bar{\sigma}}-p')
    \Gamma_{\pp}(\omega+\epsilon_{\pv'},\pv+\pv')\,,
\label{eqnSigmapp}\\
\Sigma^\sigma_{\hh}(\omega,\pv)
   =& -\int^\Lambda \dthreepp \theta(p'-k_F^{\bar{\sigma}})
    \Gamma_{\hh}(\omega+\epsilon_{\pv'},\pv+\pv')\,.
\label{eqnSigmahh}
\end{align}
In the RPA scheme, the HF term $gn_{\bar{\sigma}}$ must be removed from
$\Sigma$ since it is already contained in the HF s.p. energies
$\epsilon_{\pv}^\sigma$, but in the limit $\Lambda\to\infty$ it
vanishes anyway.

The aim of this section is the calculation of the occupation numbers
$n^\sigma_{\pv}$. In terms of the zero-temperature Green's function,
they can be obtained from \cite{FetterWalecka}
\begin{equation}
n^\sigma_{\pv} = -i\int \frac{d\omega}{2\pi} e^{i\omega\eta} G^\sigma(\omega,\pv)\,.
\end{equation}
If one keeps, as in \Eq{eqnfirstorder}, only the first-order term of the
Dyson equation, one readily obtains
\begin{multline}
n^\sigma_{\pv} = \theta(k_F^\sigma-p)\Big(1-i\int \frac{d\omega}{2\pi}
    \frac{\Sigma^\sigma(\omega,\pv)}{(\omega-\epsilon_{\pv}-i\eta)^2}\Big) \\
  -i\,\theta(p-k_F^\sigma)\int \frac{d\omega}{2\pi}
    \frac{\Sigma^\sigma(\omega,\pv)}{(\omega-\epsilon_{\pv}+i\eta)^2}\,.
\end{multline}
With the help of the residue theorem, this can be written as
\begin{multline}
n^\sigma_{\pv} = \theta(k_F^\sigma-p) \Big(1 + 
  \frac{d}{d\omega}\Sigma^\sigma_{\pp}(\omega,\pv)\Big|_{\omega=\epsilon_{\pv}}\Big)\\
  -\theta(p-k_F^\sigma)
    \frac{d}{d\omega}\Sigma^\sigma_{\hh}(\omega,\pv)\Big|_{\omega=\epsilon_{\pv}}\,.
\label{eqnn}
\end{multline}
One sees that in the case $p>k_F^\sigma$ only backward going ladders
contribute to the occupation numbers (however, remember the remark
after \Eq{eqforwardbackward}). Likewise, in the case $p<k_F^\sigma$
only forward going ladders contribute. For the numerical evaluation it
is convenient to transform the expressions for the occupation numbers
with the help of \Eqs{eqnGammapp} -- (\ref{eqnSigmahh}). For
$p>k_F^\sigma$, one gets
\begin{align}
n^\sigma_{\pv} 
  &= -\int^\Lambda \!\!\dthreek 
  \int_{-\infty}^{\OmegaF}\!\! \frac{d\omega}{\pi}
  \frac{\theta(|\kv-\pv|-k_F^{\bar{\sigma}})}
    {(\omega-\epsilon_{\pv}-\epsilon_{\kv-\pv})^2}
  \Imag\Gamma(\omega,\kv)\,.
\label{eqnnp}
\end{align} 
Note that the denominator in \Eq{eqnnp} cannot become zero because of
the upper limit of the $\omega$ integral, the theta function, and the
condition $p>k_F^\sigma$. Analogously, one obtains for $p<k_F^\sigma$
\begin{equation}
n^\sigma_{\pv} = 1+\int\!\!\dthreek 
  \int_{\OmegaF}^\infty\!\! \frac{d\omega}{\pi}
  \frac{\theta(k_F^{\bar{\sigma}}-|\kv-\pv|)}
    {(\omega-\epsilon_{\pv}-\epsilon_{\kv-\pv})^2}
  \Imag\Gamma(\omega,\kv)\,.
\label{eqnnh}
\end{equation}
In this integral, the cutoff can be omitted since the relative
momentum of the two particles is anyway limited to $|\kv/2-\pv| \leq
(k_F^\up+k_F^\down)/2$ because of the condition $p<k_F^\sigma$ and the
theta function.

In practice, as mentioned before, we replace $\Gamma$ by
$\tilde{\Gamma}$ and let the cutoff $\Lambda$ go to infinity. The
angular integrals in \Eqs{eqnnp} and (\ref{eqnnh}) can be evaluated
analytically. The integrals over $\omega$ are split into pole and
continuum contributions. The contributions of the delta functions in
$\Imag \Gamma$, see \Eq{eqnGammaspectrum}, are of course included
analytically, while the continuum contributions are computed
numerically. The remaining integrals over $k$ are done numerically,
too.

\begin{figure}
\begin{center}
\includegraphics[scale=1.2]{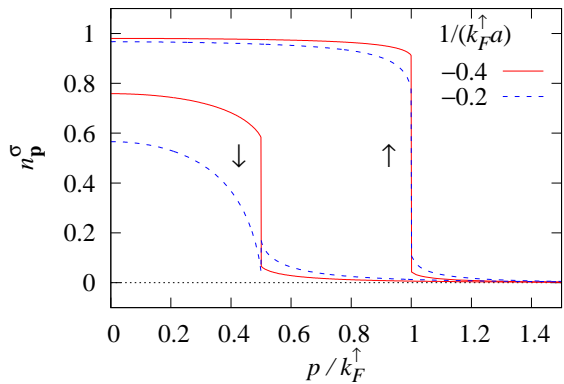}
\end{center}
\caption{Occupation numbers of majority ($\up$, upper curves) and
  minority ($\down$, lower curves) particles as function of momentum,
  for $k_F^\down = k_F^\up/2$ and two different interaction strengths
  $1/(k_F^\up a) = -0.4$ (solid lines) and $-0.2$ (dashed
  lines).\label{figplotn1}}
\end{figure}

As an example, we show in \Fig{figplotn1}
the occupation numbers for $k_F^\down = k_F^\up/2$ for two different
interaction strengths. We see that the correlations reduce
$n^\sigma_{\pv}$ below $k_F^\sigma$ and generate non-vanishing
$n^\sigma_{\pv}$ above $k_F^\sigma$. This is effect is much stronger
for the minority ($\down$) particles than for the majority ($\up$)
particles. For $1/(k_F^\up a) = -0.4$, the occupation numbers look
reasonable, but in the more strongly interacting case $1/(k_F^\up a) =
-0.2$ the jump in the $\down$ occupation numbers has the wrong
sign. If we increase the interaction further, the $\down$ occupation
numbers even become negative below the Fermi surface. This
pathological behavior is a consequence of \Eq{eqnfirstorder}, where
the correlations are treated perturbatively by truncating the Dyson
equation at first order.

Actually, from \Eq{eqnn} one sees that the jump of $n^\sigma_{\pv}$ is
given by
\begin{equation}
n^\sigma_{|\pv|\to k_F^{\sigma-}}-n^\sigma_{|\pv|\to k_F^{\sigma+}} =
  1+\frac{d}{d\omega}\Sigma^\sigma(\omega,k^\sigma_F)\Big|_{\omega=\epsilon_F^\sigma}\,.
\label{eqnjump}
\end{equation}
This has to be compared with the exact result
\begin{equation}
n^\sigma_{|\pv|\to k_F^{\sigma-}}-n^\sigma_{|\pv|\to k_F^{\sigma+}} = Z^\sigma_{k_F^\sigma}\,,
\label{eqnjumpexact}
\end{equation}
where $Z^\sigma_{\pv} = 1/[1 - d \Sigma^\sigma (\omega,\pv) / d \omega
  |_{\omega = \epsilon^{\sigma *}_{\pv}}]$ is the quasiparticle
residue of the s.p. Green's function, with the quasiparticle energy
$\epsilon^{\sigma *}_{\pv} = \epsilon^\sigma_{\pv} + \Sigma^\sigma
(\epsilon^{\sigma *}_{\pv}, \pv)$ \cite{Dickhoff}. Equations
(\ref{eqnjump}) and (\ref{eqnjumpexact}) agree to leading order in
$\Sigma^\sigma$ and the change of sign in \Eq{eqnjump} for
$d\Sigma^\sigma/d\omega < -1$ is just a consequence of the breakdown
of the expansion $1/(1-d\Sigma^\sigma/d\omega) \approx
1+d\Sigma^\sigma/d\omega+\dots$.
\subsection{Luttinger theorem}
The Luttinger theorem \cite{Luttinger} states that the correlated
occupation numbers $n^\sigma_{\pv}$ have their discontinuity still at
$k_F^\sigma = (6\pi^2 n_\sigma)^{1/3}$. In other words, if we define
\begin{align}
\delta n_{\hole}^\sigma &= \int \dthreep \theta(k_F^\sigma-p)
  (n^\sigma_{\pv}-1)\,,\label{eqndnh1}
\\
\delta n_{\particle}^\sigma &= \int \dthreep \theta(p-k_F^\sigma)
  n^\sigma_{\pv}\,,\label{eqndnp1}
\end{align}
the occupation numbers have to satisfy $\delta n_{\hole}^\sigma + \delta
n_{\particle}^\sigma = 0$.

Using the equations of the preceding subsections and the analytic
properties of the different functions in the complex plane, one can
show after some transformations that
\begin{gather}
  \delta n_{\hole}^\sigma = \int \dthreek \int \frac{d\omega}{2\pi}
  \Imag\Big(\Gamma(\omega,\kv)\frac{d}{d\omega}J_{\hh}(\omega,\kv)\Big)\,,
\label{eqndnh2}\\
\delta n_{\particle}^\sigma = \int \dthreek \int
  \frac{d\omega}{2\pi} \Imag\Big(\Gamma(\omega,\kv) \frac{d}{d\omega}
  J_{\pp}(\omega,\kv)\Big)\,.
\label{eqndnp2}
\end{gather}
In the derivation of \Eq{eqndnp2}, we made use of the cutoff
regularization, which ensures that $J_{\pp}$ falls off like $1/\omega$
for $\omega\to \infty$. Interestingly, we see that $\delta
n_{\hole}^\sigma$ and $\delta n_{\particle}^\sigma$ are independent of
$\sigma$.

In order to show that the Luttinger theorem is satisfied, we add
\Eqs{eqndnh2} and (\ref{eqndnp2}):
\begin{align}
\delta n_{\hole}^\sigma + \delta n_{\particle}^\sigma &=
   \int \dthreek \int \frac{d\omega}{2\pi}
      \Imag\Big(\Gamma\,\frac{dJ}{d\omega}\Big)
       \nonumber\\
  &= -\int \dthreek \int \frac{d\omega}{2\pi}
      \frac{d}{d\omega} \Imag\log (1-gJ)\nonumber\\
  &= 0\,.
\end{align}
Again we have used the cutoff regularization which ensures that $J\to
0$ for $\omega\to\infty$.

The proof can also be carried out with the renormalized functions
$\tilde{J}$ and $\tilde{\Gamma}$, but it is more cumbersome in that
case. If we integrate numerically our occupation numbers, which are
obtained with the renormalized functions, the Luttinger theorem is
satisfied to a precision of $\sim 10^{-4}$.

In nuclear physics, the fact that pp-RPA does not modify the sum of
particle- and hole occupation numbers in finite nuclei (having
discrete particle- and hole levels) has been known for many years, see
\Ref{VinhMau}. Recently, the pp-RPA formalism has been also applied to
the case of Bose-Fermi mixtures, and there it was also found that it
respects the Luttinger theorem for the Fermion occupation numbers
\cite{Sogo}.

\subsection{Energy density and chemical potentials}
While in the standard NSR approach the corrections to the densities
$n_\sigma$ are calculated for fixed chemical potentials $\mu_\sigma$,
we have just seen that in the zero-temperature formalism there are no
corrections to the densities due to correlations. However, there are
corrections to the chemical potentials, so that in the end the
relationships between $\mu_\sigma$ and $n_\sigma$ are changed in the
zero-temperature formalism, too. If one did a strictly perturbative
expansion, i.e., without resummation of ladder diagrams, the
difference between the relationships $n_\sigma(\mu_\up,\mu_\down)$ or,
vice versa, $\mu_\sigma(n_\up,n_\down)$ obtained in the two formalisms
should be of higher order than the expansion \cite{Dickhoff}.

Here we will calculate the chemical potentials from the energy density
$\calE$:
\begin{equation}
\mu_\sigma = \frac{\partial\calE}{\partial n_\sigma}\,.
\end{equation}
The correlation energy density, $\delta\calE = \calE-\calE_0$, where
$\calE_0 = (k_F^{\up 5}+k_F^{\down 5})/(20\pi^2 m)+gn_\up n_\down$ is
the HF energy density of the uncorrelated system, can be derived from
the following general formula \cite{FetterWalecka}
\begin{gather}
\delta\calE = -\frac{i}{2}\int_0^1 \! \frac{d\lambda}{\lambda}
  \int \!\! \dthreep
  \int \!\! \frac{d\omega}{2\pi} 
    e^{i\omega\eta}(\omega-\epsilon_{\pv}) (G_\lambda^\up+G_\lambda^\down)\,.
\label{eqnlambdaintegration}
\end{gather}
In \Eq{eqnlambdaintegration}, $G^\sigma_\lambda$ denotes the Green's
function calculated with coupling constant $\lambda g$ instead of
$g$. More precisely, since RPA theory is built on top of the HF ground
state (this is why $\calE_0$ is not the non-interacting but the HF
energy density), the coupling constant entering the HF field must not
be multiplied by $\lambda$. Anyway, this detail is not important since
the HF field vanishes in the limit $\Lambda\to \infty$.

By inserting \Eqs{eqnfirstorder} and (\ref{eqnSigmasplit}) (without
the HF term $gn_{\bar{\sigma}}$) into \Eq{eqnlambdaintegration}, one
readily obtains
\begin{multline}
\delta\calE = \frac{1}{2} \int_0^1 \frac{d\lambda}{\lambda} \int \dthreep 
  \sum_\sigma 
  \big(-\theta(p-k_F^\sigma)\Sigma_{\lambda\,\hh}(\epsilon_{\pv},\pv)\\
  +\theta(k_F^\sigma-p)
     \Sigma_{\lambda\,\pp}^\sigma(\epsilon_{\pv},\pv)\big)\,.
\end{multline}
Using the expressions given in \Secs{sectmatrix} and
\ref{secselfenergy} and exploiting the analytical properties of the
functions in the complex plane, one can show that this expression is
equal to
\begin{equation}
\delta\calE =  - \int_0^1 \! \frac{d\lambda}{\lambda}
  \int \!\! \dthreek \int_{-\infty}^{\OmegaF} \frac{d\omega}{\pi}
    \Imag \big((\Gamma_\lambda-\lambda g) J\big)\,.
\label{eqcorrelationenergy}
\end{equation}
Let us now look at the integral over $\lambda$:
\begin{gather}
\int_0^1 \frac{d\lambda}{\lambda} \Gamma_\lambda
  = \int_0^1 \! d\lambda 
    \frac{1}{\frac{1}{g}-\lambda J}
  = -\frac{1}{J}\log(1-g J)\,.
\end{gather}
Some care has to be taken in the presence of poles in $\Gamma$, since
in this case the argument of the logarithm can become negative. In
\Fig{figplotj0}, this corresponds to the region between the hh
continuum and the blue line, and between the red line and the pp
continuum. In these regions, we have $\Imag \log(1-gJ) = -\pi$. Now
\Eq{eqcorrelationenergy} becomes
\begin{equation}
\delta\calE = \int \dthreek \int_{-\infty}^{\OmegaF} \frac{d\omega}{\pi}
  \Imag \big(\log(1-gJ) + gJ\big)\,.
\label{eqnenergycutoff}
\end{equation}
Although we have assumed in our derivation that $J$ is regularized
with a cutoff $\Lambda$, we can now take the limit $\Lambda\to\infty$.
In this limit, the contribution of the last term of
\Eq{eqnenergycutoff} vanishes, and we are left with the following
compact formula for the correlation energy:
\begin{equation}
\delta\calE = \int \dthreek \int_{-\infty}^{\OmegaF} \frac{d\omega}{\pi}
  \Imag \log\Big(\tilde{J}-\frac{1}{\tilde{g}}\Big)\,.
\label{eqnenergy}
\end{equation}

In our calculation of the energy density, the integrals over $k$ and
$\omega$ are done numerically (except for the $\omega$ integral of the
pole contribution). We have not attempted to derive formulas for the
chemical potentials and we compute them by numerically differentiating
the energy density.

\begin{figure}
\begin{center}
\includegraphics[scale=1.2]{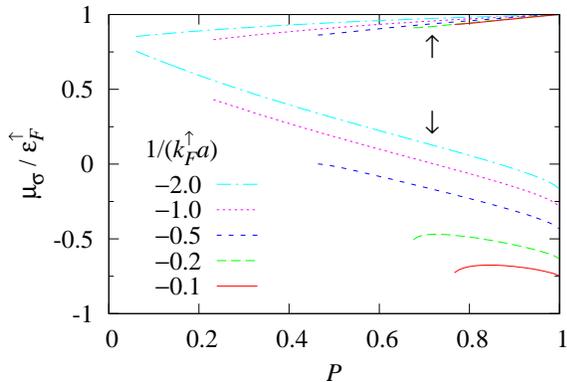}
\end{center}
\caption{Chemical potentials of the majority ($\up$, upper curves) and
  minority ($\down$, lower curves) particles as functions of the
  polarization for different interaction strengths from $1/(k_F^\up a)
  = -2$ (dash-dotted lines) to $-0.1$ (solid lines).\label{figplotmu}}
\end{figure}

In \Fig{figplotmu},
we show the chemical potentials $\mu_\up$ and $\mu_\down$ as functions
of the polarization $P$ for fixed $n_\up$ for different interaction
strengths. For each interaction strength, the polarization is varied
over the range in which the condition (\ref{conditionabovecritical})
is fulfilled, i.e., $P > P_c$. As one would expect, the chemical
potential of the minority species, $\mu_\down$, is much more strongly
lowered by interactions than that of the majority species,
$\mu_\up$. Actually, already within the ``Hartree approximation''
($\mu_\sigma = \epsilon_F^\sigma+\tilde{g} n_{\bar{\sigma}}$) it is
like that\footnote{\label{footnotehartree}As discussed before, the
  true HF shift $gn_{\bar{\sigma}}$ vanishes for the regularized
  contact interaction in the limit $\Lambda\to\infty$. However, in the
  weak-coupling limit, $\tilde{\Gamma}$ can be approximated by
  $\tilde{g}$ and this leads to a constant shift $\tilde{g}n_\sigma$
  which is usually referred to as Hartree shift (there is no exchange
  term because the interaction acts only between particles of opposite
  spin).}. At the two strongest interactions $1/(k_F^\up a) = -0.2$
and $-0.1$, one observes that $\mu_\down$ increases with decreasing
$n_\down$ at polarizations $P < 0.73$ and $P < 0.85$, respectively.

One could be tempted to say that $\partial\mu_\down/\partial n_\down <
0$ indicates an instability towards phase separation into a more and a
less polarized phase (first-order phase transition). This would be
nice, because experimentally it is found that the system separates
into a polarized and an unpolarized phase below some critical
polarization \cite{Shin}. However, the agreement would be purely
qualitative since the critical polarization observed in the experiment
is much lower than ours (e.g., in the unitary limit, it is about $0.4$
\cite{Shin}). Furthermore one should remember that at the
polarizations where we find $\partial\mu_\down/\partial n_\down < 0$,
the truncation of the Dyson equation to first order in $\Sigma$ gives
the wrong sign of the jump in the occupation numbers $n_{\pv}^\down$,
cf.~\Fig{figplotn1}. If one discards those cases where the jump of
$n_{\pv}^\down$ has the wrong sign, one should only consider
polarizations $P > 0.81$ for $1/(k_F^\up a) = -0.2$ and $P > 0.94$ for
$1/(k_F^\up a) = -0.1$, respectively.

\begin{figure}
\begin{center}
\includegraphics[scale=1.2]{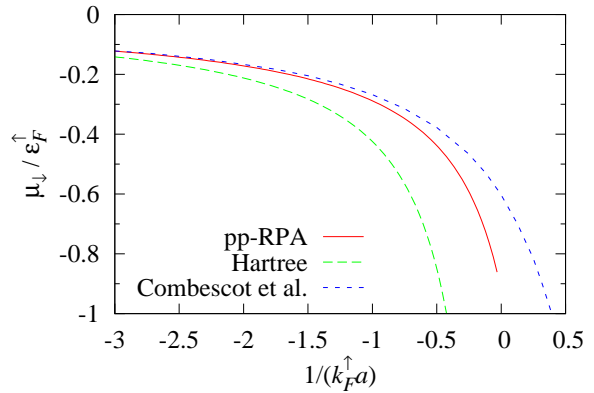}
\end{center}
\caption{Polaron chemical potential obtained within pp-RPA (solid
  line), compared with the Hartree approximation (cf. footnote)
  $\tilde{g} n_\up$ (long dashes) and with the results of
  \Ref{Combescot2007} (short dashes).\label{figpolaron}}
\end{figure}

Finally, let us discuss the limit $P\to 1$ (i.e., $n_\down \to 0$),
corresponding to the polaron. In \Fig{figpolaron},
we display our results (pp-RPA) for the polaron energy, which is equal
to $\mu_\down$ in the limit of $n_\down \to 0$, as a function of the
interaction strength. These results can be compared with the Hartree
approximation $\mu_\down = \tilde{g} n_\up$ and the results of a
calculation by Combescot et al.~\cite{Combescot2007}, which are in
very good agreement with quantum Monte-Carlo (QMC) results
\cite{Lobo,Pilati}. The calculation of \Ref{Combescot2007} is actually
also based on the T matrix, but the polaron energy is defined in a
completely different way as the pole of $G^\down$ in which the
self-energy is summed to all orders. We see that our results represent
a considerable improvement over the Hartree approximation and stay
very close to the results of \Ref{Combescot2007} up to $1/(k_F^\up a)
\sim -1$.

\section{Discussion, Conclusion, Outlook}
In this work we considered the pairing aspects of strongly polarized
Fermi gases. We worked at zero temperature within the RPA theory in
the particle-particle (pp) channel (summation of ladders) and
discussed in detail similarities and differences with the NSR
scheme. The latter gave in the recent past some pathological results
when applied to polarized Fermi gases near the unitary limit
\cite{LiuHu06,ParishMarchetti07,KashimuraWatanabe12}. These might be
related to the fact that within the NSR scheme the undressed Green's
functions building the ladder diagrams are calculated with the same
chemical potential as the final one. This is the main difference from
the zero-temperature pp-RPA formalism, where the Green's functions
depend on $k_F$ and not on $\mu$.

Using the pp-RPA, we have calculated the correction to the occupation
numbers. In particular, we have shown that the occupation numbers
satisfy the Luttinger theorem. Actually we had already shown this in a
similar scenario for interacting bosons and fermions in a mixture
\cite{Sogo}. But we have also seen that the approach breaks down when
the correlations become too strong (i.e., when the attraction becomes
too strong or the polarization becomes too small). This problem stems
from the fact that for consistency with the RPA formalism the
self-energy in the Dyson equation can only be treated perturbatively
to first order (by the way, this truncation is also made in the
original NSR scheme). It seems very unlikely that the nice properties
of the pp-RPA, such as the fact that it satisfies the Luttinger
theorem exactly, remain valid if the Dyson series is summed up (of
course the error may be quantitatively small).

We have also computed the corrections to the chemical potentials. Of
particular interest is the limit of extreme asymmetry, corresponding
to the polaron, i.e., a single $\down$ particle in a Fermi sea of
$\up$ particles. In this limit, we found good agreement with the
results by Combescot et al. \cite{Combescot2007} up to a certain
strong attraction below the unitary limit, see
\Fig{figpolaron}. Again, this is a limitation due to the truncation of
the Dyson equation at first order.

The calculations presented here were all based on the zero-temperature
formalism. It is therefore not obvious how one can generalize them to
finite temperature. As it will be shown in a separate article
\cite{PA}, it is possible to recover the results of the present work
in the $T\to 0$ limit of the finite-temperature formalism if one
includes the shift of the quasiparticle energies self-consistently
into the Green's functions $G_0$ that build the ladder diagrams and
the self-energy.

On a quantitative level, the critical polarization we obtain with our
approach is (at least in the unitary limit) much larger than the one
found experimentally \cite{Shin} and in QMC calculations
\cite{Lobo,Pilati}. This calls for an improvement of the RPA
approach. Several lines are open. One obvious drawback of the RPA is
that it calculates ground-state correlations but the ingredients to
RPA, e.g., the occupation numbers, are given by the non-correlated
ones, cf. the step functions in \Eqs{eqnJhh} and (\ref{eqnJppcutoff}).
Since we have calculated the correlated occupation numbers, a natural
idea would be to insert those in an improved RPA and iterate to
self-consistency. This would probably wash out the discontinuities in
$J$ (cf. \Fig{figplotj0}) and thereby reduce the critical
polarization. Such a procedure is often applied in nuclear physics and
called there ``renormalized RPA'' \cite{Catara,Delion}. A still
farther reaching (but numerically difficult) improvement of RPA is the
so-called ``self-consistent RPA'', in which not only occupation
numbers are included self-consistently, but also vertex corrections
\cite{Delion,Hirsch,Storozhenko}.

Another probably important thing would be the inclusion of screening
of the interaction, which is known to reduce the gap in the balanced
case (Gor'kov-Melik-Barkhudarov correction \cite{GMB}) and which
therefore could also reduce the critical polarization. However, it is
not clear how to include these particle-hole effects consistently into
the particle-particle ladders.


\begin{thebibliography}{*}
\bibitem{Eagles} D.M. Eagles, Phys. Rev. 186, \textbf{456} (1969).
\bibitem{Leggett} A.J. Leggett, J. Phys. (Paris) \textbf{41}, C7-19 (1980).
\bibitem{NSR} P. Nozi\`eres and S. Schmitt-Rink, J. Low
  Temp. Phys. \textbf{59}, 195 (1985).
\bibitem{Greiner} M. Greiner, C.A. Regal, and D.S. Jin, Nature
  \textbf{426}, 537 (2003).
\bibitem{Shin} Y. Shin, C.H. Schunck, A. Schirotzek, and
  W. Ketterle, Nature \textbf{451}, 689 (2008).
\bibitem{FF} P. Fulde and R.A. Ferrell, Phys. Rev. \textbf{135}, A550 (1964).
\bibitem{LO} A.I. Larkin and Yu.N. Ovchinnikov, Sov. Phys. JETP
  \textbf{20}, 762 (1965).
\bibitem{Punk} M. Punk, P.T. Dumitrescu, W. Zwerger, Phys. Rev. A
  \textbf{80}, 053605 (2009).
\bibitem{FetterWalecka} A.L. Fetter and J.D. Walecka, \textit{Quantum
  Theory of Many-Particle Systems} (McGraw-Hill, New York, 1971).
\bibitem{Perali2002} A. Perali, P. Pieri, G.C. Strinati, and
  C. Castellani, Phys. Rev. B \textbf{66}, 024510
\bibitem{RingSchuck} P. Ring and P. Schuck, \textit{The Nuclear
  Many-Body Problem} (Springer-Verlag, Berlin, 1980)
\bibitem{Luttinger} J.M. Luttinger, Phys. Rev. \textbf{119}, 1153 (1960).
\bibitem{LiuHu06} X.-J. Liu, H. Hu, Europhys. Lett. \textbf{75}, 364
  (2006).
\bibitem{ParishMarchetti07} M.M. Parish, F.M. Marchetti, A. Lamacraft,
  and B.D. Simons, Nature Phys. \textbf{3}, 124 (2007).
\bibitem{KashimuraWatanabe12} T. Kashimura, R. Watanabe, and
  Y. Ohashi, J. Low Temp. Phys. \textbf{171}, 355 (2013).
\bibitem{PA} P.-A. Pantel, D. Davesne, and M. Urban, in preparation.  
\bibitem{Chevy} F. Chevy and C. Mora, Rep. Prog Phys. \textbf{73}
  112401 (2010).
\bibitem{Cooper} L.N. Cooper, Phys. Rev. \textbf{104}, 1189 (1956).
\bibitem{Dickhoff} W.H. Dickhoff and D. Van Neck, \textit{Many-Body
  Theory Exposed!} (World Scientific, New Jersey, 2008).
\bibitem{VinhMau} A. Bouyssy and N. Vinh Mau, Nucl. Phys. A
  \textbf{229}, 1 (1974).
\bibitem{Sogo} T. Sogo, P. Schuck, and M. Urban, Phys. Rev. A
  \textbf{88}, 023613 (2013).
\bibitem{Combescot2007} R. Combescot, A. Recati, C. Lobo, and F. Chevy,
  Phys. Rev. Lett. \textbf{98}, 180402 (2007).
\bibitem{Lobo} C. Lobo, A. Recati, S. Giorgini, and S. Stringari,
  Phys. Rev. Lett. \textbf{97}, 200403 (2006).
\bibitem{Pilati} S. Pilati and S. Giorgini,
  Phys. Rev. Lett. \textbf{100}, 030401 (2008).
\bibitem{GMB} L.P. Gor'kov and T.K. Melik-Barkhudarov,
  J. Exp. Theor. Phys. (USSR) \textbf{40}, 1452 (1961) [translation:
    Sov. Phys. JETP \textbf{13}, 1018 (1961)].
\bibitem{Catara} F. Catara, G. Piccitto, M. Sambataro, N. Van Giai,
  Phys. Rev. B \textbf{54}, 17536 (1996).
\bibitem{Delion} D.S. Delion, P. Schuck, and J. Dukelsky, Phys. Rev. C
  \textbf{72}, 064305 (2005).
\bibitem{Hirsch} J.G. Hirsch, A. Mariano, J. Dukelsky, and P. Schuck,
  Ann. Phys. (N.Y.) \textbf{296}, 187 (2002).
\bibitem{Storozhenko} A. Storozhenko, P. Schuck, J. Dukelsky, G. R\"opke,
  and A. Vdovin, Ann. Phys. (N.Y.) \textbf{307}, 308 (2003).
\end{thebibliography}
\end{document}